\documentclass[aps,twocolumn,superscriptaddress,showpacs,floatfix]{revtex4} 
\usepackage{changebar}
\usepackage{graphicx}
\usepackage{epsfig}
\usepackage{subfigure}
\usepackage{amsmath}
\usepackage{rotating}
\usepackage{amsfonts}
\usepackage{amssymb} 
\newlength {\oldtextheight}
\newlength {\oldheadsep}
\psfigdriver{dvips}
\usepackage{changebar}
\usepackage{graphicx}
\usepackage{epsfig}
\usepackage{subfigure}
\usepackage{amsmath}
\usepackage{rotating}
\usepackage{amsfonts}
\usepackage{amssymb}  
\psfigdriver{dvips}

\begin{document}
\title{Origin of `end of aging' and sub-aging  scaling behavior in glassy dynamics}  
\author{Paolo Sibani}
\email[]{paolo.sibani@ifk.sdu.dk}
\affiliation{Institut for Fysik og Kemi, SDU, DK5230 Odense M, Denmark} 
 \author{Gregory G. Kenning}
\email[]{gregory.kenning@iup.edu} 
\affiliation{Dept. of Physics, Indiana University of Pennsylvania
Indiana, PA 15705-1001 } 
\date{\today} 
\begin{abstract} 
 Linear response functions of aging systems  
  are  routinely  interpreted using the  scaling variable $t_{\rm obs}/t_{\rm w}^\mu$,
where $t_{\rm w}$ is  the time at which the   field conjugated to the response 
is turned on or off, and where $t_{\rm obs}$ is the `observation' time elapsed from the field change. 
The response curve obtained for
different values of $t_w$ are  usually collapsed using  
values of  $\mu$  slightly below  one, a scaling behavior generally known as \emph{sub-aging}.  
  Recent spin glass Thermoremanent Magnetization experiments
  have  shown that  the value of $\mu$ is strongly 
affected by the form of the  initial cooling protocol  
(Rodriguez et al., Phys. Rev. Lett. 91, 037203, 2003), and even 
 more importantly,   (Kenning et al., Phys. Rev. Lett. 97,  057201, 2006)   
  that the     $t_{\rm w}$ dependence of the response curves 
 vanishes  altogether in  the limit $t_{\rm obs} \gg t_{\rm w}$. The latter result shows that
   the  widely used $t_{\rm obs}/t_{\rm w}^\mu$ scaling of 
linear response data cannot be generally valid, and casts some doubt on the
theoretical significance of the exponent $\mu$.
In this work,  a common  mechanism is proposed for  the origin  
of both  sub-aging and end of aging behavior in glassy dynamics. The mechanism  combines real  and 
configuration space
properties of the  state produced by the initial thermal quench which initiates the aging process. 
\end{abstract}  
 
\maketitle
\section{Introduction}  
 In  glassy systems, a  thermal quench initiates a so called  aging process,  
 whereby  physical observables, e.g. the energy,  
 slowly change as a function of  `age', the term conventionally denoting 
 the  time $t$ elapsed from  the   quench. 
In  non-stationary 
processes, conjugated linear response  and correlations functions   generally
 depend on two time arguments,  e.g. the  system age $t$, and its value 
 $t_{\rm w}$ at the moment where the  field is switched on,  or off.
 However,   both  functions appear to actually depend 
 on  a single  scaling variable, namely
  $t_{\rm obs}/t_{\rm w}^\mu$, where
$t_{\rm obs}\stackrel{def}{=} t -t_{\rm w}$~\cite{Struik78,Alba86,Ranjith08}.  
The `observation'  time $t_{\rm obs}$,   
(widely   denoted by the symbol $t$ in the literature)  is often used  in lieu of the system age
 as an independent time variable. The exponent $\mu$ is generally near one, 
and the terms \emph{super-aging},   \emph{sub-aging} and \emph{pure} or \emph{full} aging 
are used to discriminate between  cases with  $\mu >1$,  $\mu <1$ and $\mu=1$, respectively.  
Sub-aging (henceforth SA)  has mainly  been observed  in spin-glass  Thermoremanent Magnetization 
(TRM) data~\cite{Alba86,Zotev03,Rodriguez03,Parker06}, usually after subtracting  a `stationary'
term which describes the response immediately after the field is cut.

As a  quench is unavoidably   carried out  at a finite cooling rate, 
 the point  $t=0$ on the time axis    eludes a sharp   definition  in  an  experimental 
setting.  Possibly as a consequence thereof, the  scaling form of response functions is,
in spite  of a protracted 
debate~\cite{Rinn00,Berthier02,Suzuki03,Zotev03,Rodriguez03,Sibani06a,Kenning06},
only partially  understood.
  E.g.  the physical significance of $\mu$ (and of the additional time
scale it brings along)  is unclear,  even more so in the light 
of the  recent discovery  that  the TRM     loses
its  $t_{\rm w}$ dependence  in the limit  $t_{\rm obs}/t_{\rm w} \rightarrow \infty$~\cite{Kenning06}.
 This  observation   falsifies the long-held hypothesis that SA could be a scaling form  generally 
  applicable to  linear response functions.    Coined in Ref.~\cite{Kenning06},
   the term `end of aging' (EOA)   is also adopted here. It describes  that, for large values
   of $t/t_w$, the $t_w$  dependence of the TRM disappears and is replaced 
 a simple  logarithmic time decay.  The term  should not be construed as   contradicting   the   fundamental
unity  of aging dynamics   which subtends   our  description.  

 In this work,  an expression for the  TRM decay, Eq.~\eqref{gen_res},  is derived   which, 
 depending on  the value of the  model parameter $x$, interpolates between  
 pure aging (PA), SA  and EOA  behavior.
Our derivation  combines    well known  configuration  
and real space properties of spin glasses and supplements them with a model 
assumption regarding the spatial  heterogeneity of the system configuration
at the beginning of the isothermal aging process. Our   analytical expression  for the TRM decay 
 is derived by   averaging the magnetic response of independent  domains over a
 suitable distribution characterizing their  initial state.  
Even though  the ratio $t_{\rm obs}/t_{\rm w}^\mu $ nowhere  appears in the treatment,
  the    TRM decay curves produced by our Eq.~\eqref{gen_res} can,  in accordance with standard  practice,
   be  \emph{empirically collapsed}  in the  SA regime  using 
$t_{\rm obs}/t_{\rm w}^\mu $  as a  scaling   variable. 
\section{Linear response and domain heterogeneity}  
Thermalized domains whose linear size grows with  the   thermal correlation length~\cite{Kisker96}
are  ubiquitous  in aging systems with short range interactions.
Real space scaling  descriptions~\cite{Bray87}  rely on their properties
to  account for several dynamical  properties, e.g. so-called 
`chaos' effects.  In a physics
 context, hierarchies were  brought into focus by 
the  Parisi equilibrium solution of   a mean field model~\cite{Parisi83}, 
a solution which has inspired hierarchical models of spin-glass  dynamics, 
e.g.~\cite{Joh96}. However, 
the hierarchical nature of relaxation dynamics in complex
system  is a  generic  feature~\cite{Simon62} which  complements  
 real space descriptions and which 
  has e.~g. been   confirmed  numerically 
 by   solving  master equations in 
 fully enumerated sets of states~\cite{Sibani93,Sibani94}.  

Hierarchical  models, see  e.g. Refs.~\cite{Grossmann85,Sibani89,Hoffmann97},  
  assume that nested ergodic components~\cite{Palmer82}   exist  in  
 configuration space.  In simple cases, the  solution of  the pertinent  master equations  
shows  PA   behavior~\cite{Sibani89}. For thermally activated
dynamics, this result can be heuristically explained  as follows: 
 Any  ergodic component  of the  hierarchy is at any time $t$ indexed by a real valued  `dynamical barrier' $b(t)$.
On  the (Arrhenius)  time  scale $t_b(t) = C \exp(b(t)/T)$ the component is 
 near a state of internal thermal equilibrium. The constant $C$ is the
 \emph{fluctuation time}, i.~e. the  smallest  relevant 
time scale of the dynamics.  
Consider now a system  starting 
 in a  component  with vanishing initial  barrier $b(t=0) \approx 0$.
  After  aging isothermally  for a time $t=t_w$, the component  
 is   characterized  by a barrier $b(t_w)\propto  T \ln(t_w/C)$.   
The   dynamics on scales   $t_{\rm obs}<t_w$ has character of quasi-equilibrium fluctuations
within  the same component, while off-equilibrium processes involving larger components take over  for $t_{\rm obs} > t_w$.
The age  $t_w$ hence separates the two dynamical regimes observed for  aging
systems, and constitutes  the  dominant  time scale for $t>t_w$. From this observation,  
pure aging heuristically follows.  
A mathematically more precise route leading to the same conclusion
relies on the fact that, within a   
 hierarchy indexed by energy barriers,  only 
thermal energy fluctuations of \emph{record} magnitude   are able to   trigger    
 irreversible changes of ergodic component, or \emph{quakes}.  
Since the  statistics of record-sized fluctuations in a stationary series is known
 analytically~\cite{Sibani93a,Sibani03},  assuming 
 that all physical   changes  are statistically subordinated
 to the quakes leads to   \emph{record dynamics}~\cite{Sibani03,Sibani05,Fischer08} 
and to analytical formulas,   such as 
Eq.~\ref{M_0},   for   one and two-point averages of aging processes.
In spite of their simplicity,
the  above  ideas    rationalize a large amount of experimental 
evidence,~\cite{Vincent91,Jonason98,Sibani06a}. Yet,  they neither 
account for SA  nor for EOA scaling behaviors. In order to  do so,
  the spatial and temporal  heterogeneity of spin
glasses,  experimentally  demonstrated  by Chamberlin~\cite{Chamberlin99},
must be properly taken into account.

In the present model, independently relaxing   domains 
of a glassy  system are  all  endowed with  the same type
of  hierarchically structured 
 configuration space.
Nonetheless, their  respective  contributions  to the overall linear response  
are different (albeit related)  functions of time. The assumption is 
that domains  find themselves in states  characterized by 
 different dynamical barriers at the end of the initial quench,  or
 equivalently, at  the beginning of the isothermal aging process.

The physical mechanism behind the difference is likely  related to the
 way in which the cooling process proceeds near the glass transition temperature,
 see e.~g. ref.~\cite{Parker06,Kenning09}. 
Here, the spatial heterogeneity is   heuristically  described 
  by a distribution  of initial barriers $P(b)$. 
We checked that a flat distribution supported in the  interval $(0,b_M^*)$
and an exponential distribution with average $b_M^*$ lead to  similar
behaviors. The existence
of a finite first moment  of $P(b)$ seems to be the crucial feature. 
Since the exponential form leads to simpler  closed form  expressions,
this form  is chosen for    mathematical convenience. 

In summary the initial state of the 
 aging process feature domains described by the  dynamical barriers    distribution   
\begin{equation}
 P(b) = \frac{1}{b_M^* }\exp(-b/b_M^*).
 \label{exp_distr}
 \end{equation}
 Let $t_M^*$ be the Arrhenius time associated to 
 $b_M^*$. 
As we shall see, the quantity 
\begin{equation}
 x  \stackrel{\rm def}{=} \frac{T}{b_M^*}=\frac{1}{\ln(t_M^*/C)} 
 \label{x_def}
 \end{equation}
  controls  all 
deviations from pure aging behavior.
 Note that $x >0$, that  $x\rightarrow 0$ for $t_M^* \rightarrow \infty$, and that
$x\rightarrow \infty$ for $t_M^* \downarrow C$.    
  
\section{Pure aging  approximation of TRM data}   
The additive  contribution  to the  magnetic response of a single  domain  is 
 in the present theory given  by  Eq.~\eqref{dom_mag}. In this espression   
the  function $M_0(t/t_{\rm w})$  describes the TRM of a domain 
initially in a  state with vanishing  dynamical barrier.   

The functional form chosen for  $M_0$ reflects  that  record dynamics is a homogeneous 
stochastic process in the single `time' variable $\log(t/t_{\rm w})$. By standard arguments, 
all moments of the process, including the 
average response,  admit eigenvalue expansions where  $\log(t/t_{\rm w})$ replaces
time. The generic term 
in  such expansions is proportional to $(t/t_{\rm w})^{\lambda_k}$, where $\lambda_k$ is the $k$'th 
 relaxation eigenvalue. In practice, the  expansion can be cut after few terms and, as shown graphically in the
 Appendix, two terms (one term less than in Ref.~\cite{Sibani06a}) already  provide an acceptable parameterization 
 of the TRM decay in the  PA approximation.  

Summarizing, the pure aging scaling ansatz
for the TRM can be written as
\begin{equation}
M_0(z) = 
M_{\rm I} + \eta(z-1)  
  \sum_{k=1}^2 \frac{a_k}{\lambda_k}
 \left(
 z^{\lambda_k} -1
\right),  
\label{M_0}
\end{equation} 
where $z = t/t_{\rm w}$, and where $\eta$ is the Heaviside step function, and  
 $M_{\rm I}$ is  the `initial' value of the TRM, which for simplicity 
 is  treated  as a   parameter.
 According to the formula,   the TRM  remains constant and equal to $M_{\rm I}$ until  the 
magnetic field is cut.  
 The notation $a_k/\lambda_k$ for the prefactor 
 is chosen to simplify the form of the   rate of magnetization change  
\begin{equation}
r_{\rm TRM,0}(t,t_{\rm w}) = \frac{1}{t} 
 \sum_{k=1}^2
a_k \left( \frac{t}{t_{\rm w}}\right)^{\lambda_k} .
\label{main_fit_formula}
\end{equation}
The (negative) pre-factors and exponents  are  for completeness tabulated  in 
the Appendix.  

\section{Origin of sub-aging and end of aging}   
Let  $T$ denote  the isothermal aging temperature. As mentioned, $C$ denotes  the smallest
relevant relaxation time, i.e. the time associated to the smallest
energy barrier in the energy landscape of a single domain. 
Consider now a domain characterized by 
the  initial barrier $b^*$, or  equivalently,  by the  Arrhenius time  
$t^*=C \exp(b^*/T)$. If $t_{\rm w} < t^*$, the  behavior at time $t=t_{\rm w}$ remains  controlled by  
 the initial barrier $b^*$ and the   domain's contribution  to the TRM 
   correspondingly depends on   $t/t^*$. 
 Conversely, if   $t_{\rm w}  > t^*$ 
the initial barrier has  been surmounted at $t_{\rm w}$, 
 and the scaling variable  is   hence $t/t_{\rm w}$.   
In real space,  the size of the domain   grows as
a function of the dynamical barrier $b(t)$.
E.~g. a  power-law growth in the  time domain~\cite{Kisker96} corresponds to an exponential
growth  in  $b(t)$. In our case,  
$b(t)=\max(T\ln(t/C), b^*)$, and  domain   growth is delayed 
up to the Arrhenius time $t^*$, compared to the case  $b^* = 0$.

Returning   to the form of the response, the contribution   of a domain with  initial dynamical  barrier $b^*$  is  
 \begin{equation}
 m(t,t_{\rm w},b^*) = \eta(t_{\rm w}-t^*) M_0(t/t_{\rm w} ) + \eta(t^*-  t_{\rm w}) M_0(t/t^* ),
\label{dom_mag}
\end{equation} 
where $\eta$ is again the Heaviside function, where $M_0$ is given in Eq.~\eqref{M_0} and where  $t \ge t_{\rm w}$.
The formula embodies the key feature  of hierarchical   relaxation
 without  reference  to any specific  model. Secondly, it 
introduces spatial heterogeneity, as  $b^*$, or  equivalently, the   Arrhenius time $t^*$,
is allowed to differ across the domains. 
At $t_{\rm w} = t^*$,   the Heaviside function  $\eta$   switches  between the two  
scaling forms available for  the magnetic response of a single domain.  If  the   barriers of the  different
domains in the system  are all initially near zero,  only the 
first term contributes. and  the  pure aging behavior given by 
$M_0(t/t_{\rm w})$  goes through at the macroscopic level.
In the general case, the TRM decay takes the form   
\begin{eqnarray} 
M(t,t_{\rm w}) &=& M_0(t/t_{\rm w}) \int_0^{T\ln(t_{\rm w}/C)} P(b)db + \label{bas_form}\\
 \int_{T\ln(t_{\rm w}/C)}^{T\ln(t/C)} 
&M_0&(\frac{t}{C}e^{-b/T})P(b)db + 
M_{\rm I} \int_{T\ln(t/C)}^{\infty} P(b)db,\nonumber 
\end{eqnarray}  
where  $P(b)$ is the   probability density for a domain with barrier  in the initial state.

\begin{figure*} 
$ 
\begin{array}{ccc} 
\includegraphics[width=0.31\linewidth,height= 0.31\linewidth]{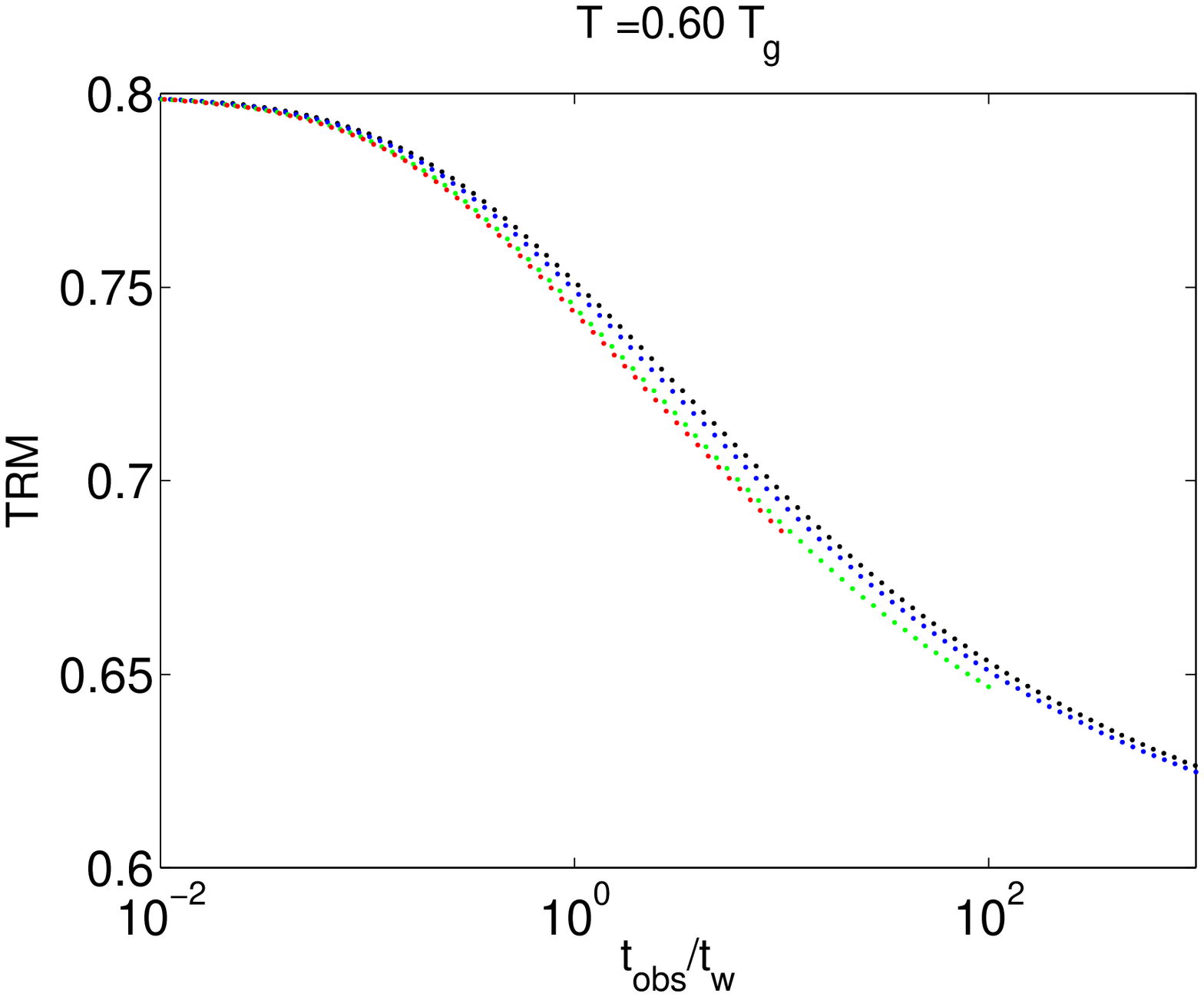} &
\includegraphics[width=0.31\linewidth,height= 0.31\linewidth]{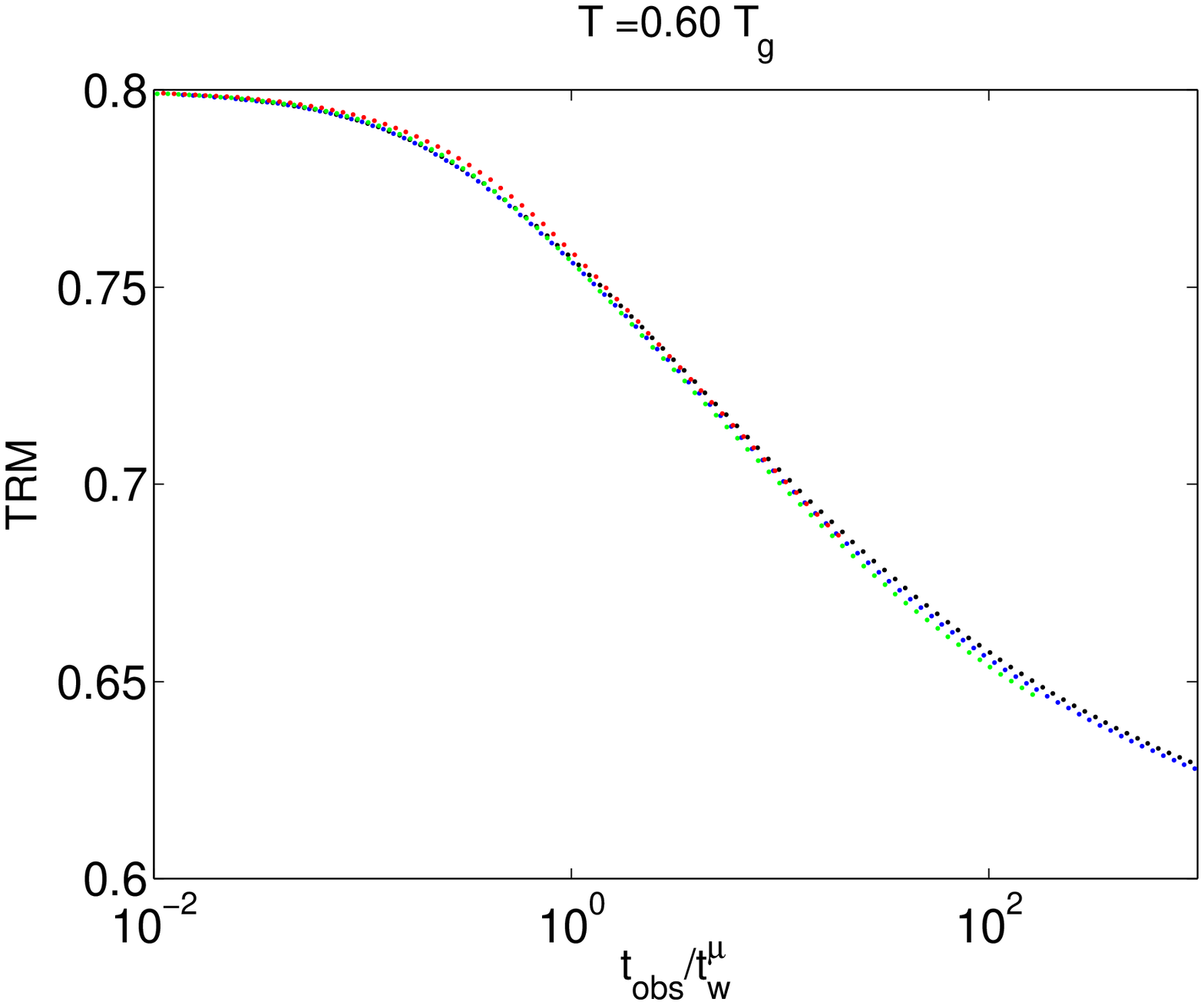}  &
\includegraphics[width=0.31\linewidth,height= 0.31\linewidth]{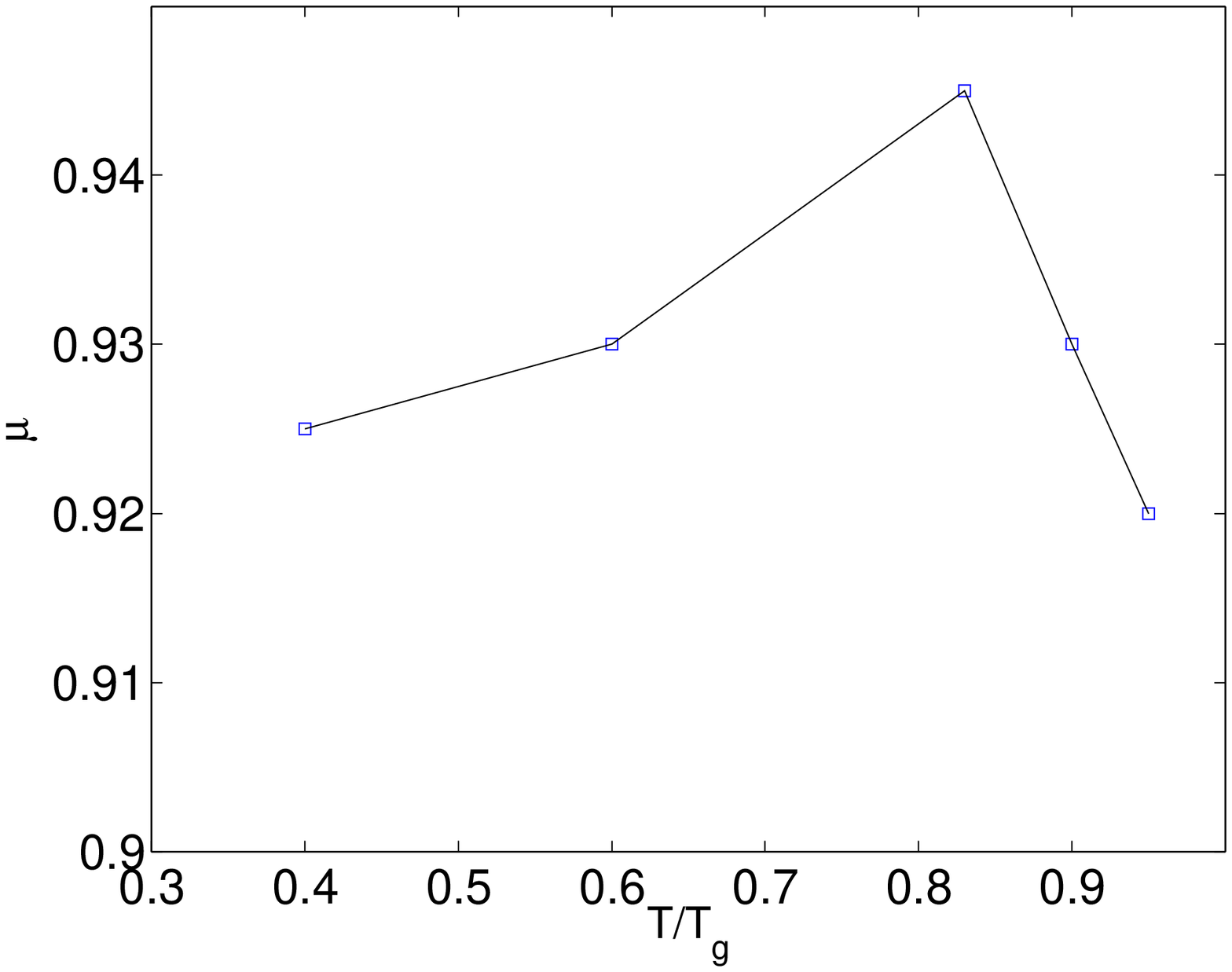} \\ 
\end{array} 
$  
\caption{ (Color on line)  The Thermoremanent Magnetization calculated  
according to Eq.~\eqref{expo_M_2} is plotted for $t_{\rm w} =50,100,1000$ and $10000$ (black, blue,
green and red) and for 
$T=0.60 T_{\rm g}$  versus $t_{\rm obs}/t_{\rm w}$, (left figure),
and versus $t_{\rm obs}/t_{\rm w}^\mu$, (middle). The left figure shows  that  sub-aging  is present, and 
the middle figure that  the standard scaling procedure  collapses the data. 
Scaling plots for  other temperatures are also performed (not shown). In each case 
a $\mu$ value is estimated as the one providing the best collapse.  
In the rightmost figure, the empirical values of $\mu$   are plotted versus $T/T_{\rm g}$,
showing a clear maximum at $T/T_{\rm g} = 0.83$. The line is only a guide to the eye.
}
\label{subagingplots}
\end{figure*} 
  
Inserting  Eq.~\eqref{exp_distr} into Eq.~\eqref{bas_form}, and 
using simple algebraic manipulations, one arrives at 
  \begin{eqnarray}
 M(t,t_{\rm w}) &= &M_{\rm I} \left( \frac{t}{C}\right)^{-x}+ 
  \left( 1 -\left( \frac{t_{\rm w}}{C}\right)^{-x} \right) M_0(t/t_{\rm w}) \nonumber \\
 &+& x \left( \frac{t}{C}\right)^{-x} \int_{1}^{t/t{\rm w}} M_0(z) z^{x-1} dz.
 \end{eqnarray}  
Using  the parameterization of $M_0$ given in  Eq.~\eqref{M_0},   the last expression  becomes
 \begin{eqnarray}
  \label{expo_M_2}
&\left. \right.&M(t,t_{\rm w}) = M_{0}(t/t_{\rm w}) \nonumber \\
 &+ & \left( \frac{t_{\rm w}}{C}\right)^{-x}
 \sum_{i=1}^2 \frac{a_i}{\lambda_i +x}   
 \left[ 
 \left( \frac{t}{t_{\rm w}}\right)^{-x} - \left( \frac{t}{t_{\rm w}}\right)^{\lambda_i}  
  \right].
  \label{gen_res} 
 \end{eqnarray} 
 Pure aging.is achieved in  the limit $x=\infty$, i.e. when all initial barriers are
equal to zero with probability one.  
   Sub-aging is present for  \emph{intermediate} values of $x$.
   Let now $\lambda_2$ be  the largest 
of the two  decay exponents characterizing  the pure aging regime.
  If and only if  $x < -\lambda_2$,  the asymptotically dominant contribution 
  to the TRM for large $t$ is proportional to $\left( \frac{t}{C}\right)^{-x}$.
  To bear this  out, we first re-write   our last equation  as
  \begin{eqnarray} 
 &\left. \right.& M(t,t_{\rm w}) = M_{0}(t/t_{\rm w}) \nonumber \\
 &+ & \left( \frac{t}{C}\right)^{-x}
 \sum_{i=1}^2 \frac{a_i}{\lambda_i +x} 
 \left( 1 -   \left( \frac{t}{t_{\rm w}}\right)^{\lambda_i +x}  
  \right).
  \label{gen_res_2} 
 \end{eqnarray}  
The  pure aging term $M_0(t/t_w)$  decays the fastest and can be neglected. In what remains,
 the $t_w$  dependent term   having 
the  slowest decay   is $\left( t/t_w \right)^{\lambda_2 + x}$.
In order for the EOA behavior to set in,  this term must be much
smaller than one.
E.g. a relative deviation of the TRM curve from EOA   equal to $1/10$ 
  is reached  at time   
\begin{equation} 
t_{\rm EOA} =  \left( 10 \right)^{\frac{1}{|x+\lambda_2|}} t_w.
 \label{expo_M_3}
 \end{equation}
Since  ${\frac{1}{|x+\lambda_2|}}$ is very large when $x + \lambda_2 \approx 0$,  
 the  model predicts that EOA may occur  on a time scale which  diverges very rapidly with $t_w$.
 This is qualitatively in  accord with the experimental observations   of Ref.~\cite{Kenning06}. 
Secondly, when the  exponent  $x$ is numerically small,  
the expansion  $(t/C)^{-x} = 1 -x \ln(t/C) + \mathcal{O}((x\ln(t/C))^2)$ is applicable, 
 and the  TRM decays  in a nearly  logarithmic fashion  for a wide range of $t$, 
  also in accord with  the experimental findings.

\section{On $ t_{\rm obs}/t_w^\mu$  scaling}  
Equation~\eqref{gen_res} features a clear sub-aging behavior  
without any reference to   the widely used scaling variable $t/t_w^\mu$.
Nevertheless,   model generated TRM data can be 
  empirically scaled in the traditional manner
  for intermediate values of $t$ and $t_w$. 
  This is checked   numerically  
\emph{(i)} by evaluating  Eq.~\eqref{gen_res}, with  
   $C=1$, $t_{\rm M}^* = 10$ and with all other parameters  given in
   Table.~\ref{num_values}, and \emph{(ii)} by scaling the curves obtained as usually
   done for experimental data. 
The left panel in Figure~\eqref{subagingplots} only confirms that  $t_{\rm obs}/t_{\rm w}$ 
scaling is not satisfactory for the parameter values utilized. The four curves  plotted all  pertain to $T=0.6 T_g$.
The values of $t_w$ used are  $t_w=50,100,1000$ and $10000$.
The exact same data are   shown in the second panel of the same figure. They are now  
 plotted versus
$t_{\rm obs}/t_w^\mu$, with the value of   $\mu$  chosen as the one  judged to produce the best 
data collapse.  
 Additional  curves (not shown) were  similarly obtained by evaluating Eq.~\eqref{gen_res}
 for $T/T_{\rm g}=0.4, 0.6, 0.9,$ and $T=0.93$.
 The empirical   $\mu$ values thus obtained gauge how close the TRM is to
 pure aging. These values  are plotted versus $T/T_{\rm g}$ in the 
  right panel of the same figure (the line is only a guide to the eye).  
Interestingly,   $\mu$ versus $T$ peaks   at $T/T_{\rm g}= 0.83$, the very temperature  
where  $\mu$   is  experimentally   closest to one~\cite{Rodriguez03}.  
 
In summary,     Eq.~\eqref{gen_res}  fully contains  the standard subaging behavior 
 widely seen in spin glasses. Furthermore, it  demonstrates that the applicability 
 of $t_{\rm obs }/t_w^\mu$ scaling   does not \emph{per se}   endow
 the exponent $\mu$ with  physical significance, which  is plainly absent in our case. 

 \section{Summary and Outlook}  
In this work, the known  scaling properties
of off-equilibrium linear response functions in 
spin glasses have been  accounted for by  combining
two aspects of complex dynamics: the hierarchical 
relaxation of independently thermalizing domains, coupled   
with the spatial heterogeneity of the initial domain configurations, as defined by 
their initial  dynamical barriers. These barriers would   uniformly vanish
for a system conforming to pure aging scaling behavior.  Our analysis 
 relies on generic properties of complex dynamics, and should therefore 
 be  widely applicable to glassy systems with short range interactions.
 These might include quantum  spin glasses, whose critical behavior   has recently been  
 investigated~\cite{Jonsson07,Ancona08},
and irrespective of whether a    true equilibrium phase transition exists~\cite{Ancona08}
  or not~\cite{Jonsson07}.  

The  distribution of  initial dynamical barriers plays a pivotal role in the
theory. Arguably, its form  depends on the cooling protocol, e.g. 
fast cooling could give a distribution more sharply
 peaked at zero, and lead to a  relaxation scaling form closer  to  pure aging.
  The width of the initial barrier distribution  is expressed by the exponent $x$, which
is \emph{experimentally accessible} as the logarithmic slope of  the TRM decay 
for very large values of $t/t_w$, i.e. in the dynamical  regime where the
$t_w$ dependence of  the data is absent.   It should  therefore  be possible to 
empirically study, via $x$, 
how the initial cooling protocol
affects the distribution of initial barriers and, indirectly, the subsequent relaxation dynamics.
\newpage
 \section{Appendix}  
  \begin{table}
\begin{tabular}{|l|l|l|l|r|} 
   \hline 
$T/T_g$ & $a_1$ & $\lambda_1$ & $a_2$ & $\lambda_2$\\ \hline
$0.40$ &$-0.0878$ & $-2.7575$ & $-0.0289 $ & $-0.1869 $\\
$0.60$ & $-0.0981$ & $-3.0945$& $-0.0462 $ & $-0.2609 $\\
$0.83$ &$-0.1365$ & $-2.9016$& $-0.0527$ & $-0.3102 $\\
$0.90$ &$-0.1309$ & $-3.2785$& $-0.0418$ & $-0.3098 $\\
$0.95$ &$-0.1117$ & $-3.4737$& $-0.0296$ & $-0.3495 $ \\  
\hline 
\end{tabular} 
\caption{ The  first column  contains the  ratios of the 
isothermal aging temperature $T$ at which the measurements are taken 
to the critical temperature $T_g$.
The other  columns contain the pre-factors and exponents of the
two power-law terms appearing  in Eq.~\eqref{M_0}. All values are obtained by fits (not shown)
of quality similar to    Fig,~\eqref{big_rate}.}
\label{num_values}
\end{table}  
We describe in this Appendix  how the parameter values  entering  
 Eq.\ref{M_0} are estimated by fitting to experimental TRM data.
  The   data   are obtained according to a standard procedure:
 a Cu$_{0.94}$Mn$_{0.06}$ spin glass sample
 is  rapidly   quenched    to a temperature $T< T_g$ 
 in the presence of  a small magnetic field.   The  field is 
cut at  time $t = t_{\rm w}$, and the 
  magnetization decay  is then recorded~\cite{Alba86,Zotev03,Rodriguez03,Parker06}
for $t>t_{\rm w}$. 

The parameters  shown in  Table~(1) are used 
\emph{empirically determine}, on the basis
of Eq.~\ref{gen_res}, the SA exponent $\mu(T)$. The small deviation of the experimental data  
from the pure aging form given in  Eq.\ref{M_0} implies, of course, a small systematic error.  

Of special importance   are
the values of the dominant exponent $\lambda_2$ which are listed in the last column of the
table: According to  Eq.~\ref{expo_M_3}, the  quantity  $\lambda_2 + x$ determines the time scale for the 
onset of $EOA$ behavior.

\begin{figure}[t]
    \centering 
   \includegraphics[width=\linewidth]{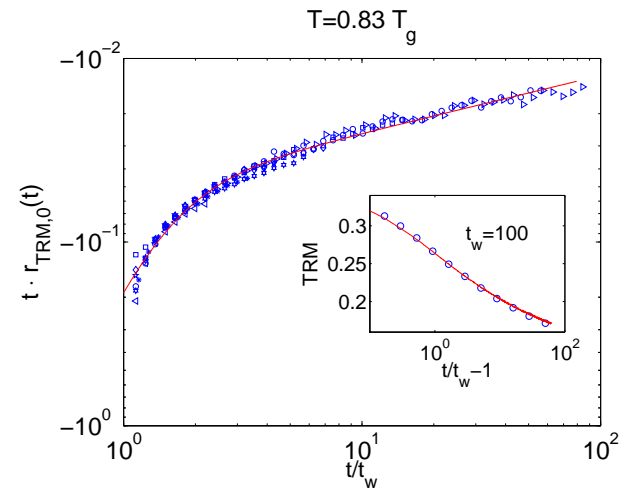}  
   \caption{ (Color on line) 
The   TRM decay rate, multiplied by the age
  $t$ is   plotted versus $t/t_{\rm w}$
for  $T/T_{\rm g} =0.83$ and 
$t_{\rm w}= 50$ (right pointing triangles), 
$100$ (circles), $300$ (squares), $630$ (diamonds), $1000$ (pentagrams), 
$3600$ (hexagrams), $6310$ (asterisks)  and $(10000)$ (left pointing triangles). 
   The  line is  given by    Eq.~\eqref{main_fit_formula}.
The insert  compares  the    TRM decay measured at  $t_{\rm w}=100$s (red line),  
with  the theoretical estimates (blue circles) obtained  by   Eq.~\eqref{M_0}. 
}
\label{big_rate}  
  \end{figure}

Figure~\ref{big_rate} is scaling plot of the rate of (de-) magnetization multiplied by
 the system age, versus the scaling variable $t/t_{\rm w}$. At the aging  temperature $T=0.83$
 the empirical sub-aging exponent $\mu$ is very close to one~\cite{Rodriguez03,Zotev03} 
 and the data conform reasonably well 
to a PA ansatz. The approximation is worse, but still usable,
at  other aging temperatures and all  data can be 
 fit reasonably well by the full aging formula~\ref{M_0}.  
   \bibliographystyle{unsrt}
\bibliography{SD-meld}
\end{document}